\documentclass[a4paper,11pt]{article}

\usepackage{graphicx}
\graphicspath{{./figures/}}
\usepackage {amsmath,amsfonts,amssymb}
\usepackage{hyperref}
\usepackage{url}
\usepackage{subfig}

\newcommand{\R}{\mathbb{R}}
\newcommand{\C}{\mathbb{C}}

\begin{document}


\title{An implicit symplectic solver for high-precision long term integrations of the Solar System
}

\author{M. Anto\~nana   \and E. Alberdi \and J. Makazaga  \and A. Murua 
}



\date{
 University of the Basque Country (UPV/EHU) \\
 \today }

\maketitle
\begin{abstract}
We present  FCIRK16, a  16th-order implicit symplectic integrator for long-term high precision Solar System simulations.  Our integrator  takes advantage of the near-Keplerian motion of the planets around the Sun  by alternating Keplerian motions with corrections accounting for the planetary interactions. Compared to other symplectic integrators (the Wisdom and Holman map and its higher order generalizations) that also take advantage of the hierarchical nature of the motion of the planets around the central star, our methods require solving implicit equations at each time-step.  We claim that, despite this disadvantage, FCIRK16 is more efficient than explicit symplectic integrators for high precision simulations thanks to:  (i) its high order of precision, (ii) its easy parallelization, and (iii) its efficient mixed-precision implementation which reduces the effect of round-off errors. In addition, unlike typical explicit symplectic integrators for near Keplerian problems, FCIRK16 is able to integrate problems with arbitrary perturbations (non necessarily split as a sum of integrable parts).
We present a novel analysis of the effect of close encounters in the leading term of the local discretization errors of our integrator. Based on that analysis, a mechanism to detect and refine integration steps that involve close encounters is incorporated in our code. That mechanism allows FCIRK16 to accurately resolve close encounters of arbitrary bodies.  We illustrate our treatment of close encounters with the application of FCIRK16 to a point mass Newtonian 15-body model of the Solar System (with the Sun, the eight planets, Pluto, and five main asteroids) and a 16-body model treating the Moon as a separate body. We also present some numerical comparisons of FCIRK16 with a state-of-the-art high order explicit symplectic scheme for 16-body model that demonstrate the superiority of our integrator when very high precision is required.
\end{abstract}

\section{Introduction}

High precision long-term dynamic simulation of planetary systems requires computationally expensive numerical integrations. 
Efficient computation of  long-term ephemerides of the Solar System can be achieved by exploiting 
the near Keplerian motion of the planets around the Sun. The main idea is to alternate Keplerian motions with appropriate corrections accounting for the planetary interactions. This is the case of symplectic splitting integrators such as the Wisdom and Holman method~\cite{WH1991} and its higher order generalizations~\cite{Wisdom1996,Laskar2011,Blanes2013,Farres2013,Mikkola2000}.

For very long-term integrations, the propagation of roundoff errors is an important limiting factor of the final accuracy, and the standard 64-bit double precision arithmetic can be insufficient in some cases. An alternative that is used in some long-term integrations of the Solar System~\cite{Laskar2011} is provided by the 80-bit extended precision arithmetic (one bit for the sign, 15 bits for the exponent, and 64 bits for the significand).
Compared to double precision arithmetic (having 53-bit significands) 11 additional binary digits of precision are gained by making all the computations in extended precision arithmetic at the cost of approximately doubling the computing time. For further reduction of roundoff errors, one could perform all the computations in quadruple precision at the cost of a drastic increase (by a factor of twenty or more) of computing time.  A more efficient approach to gain some additional digits of precision (although not providing the full accuracy of quadruple precision,)  is to use a mixed precision methodology~\cite{Baboulin2009}.  That is, to use quadruple precision for the most critical computations and perform the rest of the computation in a lower precision arithmetic. 
In the case of explicit splitting methods, a natural mixed precision strategy might be to compute the Keplerian flows in quadruple precision, and the corrections corresponding to the planetary interactions (which are typically of smaller magnitude) in 80-bit arithmetic. This would allow us to gain a few binary digits of precision, although with a considerable increase in computing time.

Further improvement of the efficiency of numerical simulations demands the development of fast and accurate algorithms that take advantage of parallel computer architectures. Although some attempts have been made in this sense for symplectic splitting integrators for Solar System simulations~\cite{Laskar2011parallel}, the sequential nature of that kind of schemes makes it difficult to get substantial improvements from parallelization strategies.

We present FCIRK16, a code for long-term high precision integrations of systems with Hamiltonian function of the form 
\begin{equation}
\label{eq:H}
\begin{split}
H(q,p) &= H_K(q,p) + H_I(q,p),\\
H_K(q,p) &= \sum_{i=1}^{n} \left(
\frac{\|p_i\|^2}{2\mu_i} - \frac{\mu_i\, k_i}{\|q_i\|}
\right),
\end{split}
\end{equation}
where $q = (q_1,\ldots,q_n)$, $p = (p_1,\ldots,p_n)$, $q_i, p_i \in \R^3$, $i=1,\ldots,n$. The Hamiltonian $H_K(q,p)$ is the sum of $n$ uncoupled spatial Kepler problems. For each $i=1,\ldots,n$,  $\mu_i$ (resp. $k_i$) is the effective mass (resp. the gravitational constant) corresponding to the $i$th Keplerian problem.  The interaction Hamiltonian $H_I(q,p)$ is seen as a perturbation of the Keplerian Hamiltonian $H_K(q,p)$.

Our code implements a 16th order implicit symplectic method within the class of FCIRK (flow-composed implicit Runge-Kutta) schemes proposed in~\cite{Anto2018}.  FCIRK methods are similar to symplectic splitting schemes in that Keplerian motions are alternated with corrections of smaller magnitude. But while in symplectic splitting integrators the corrections are computed as the solution operator of the interaction Hamiltonian,
in FCIRK methods, such corrections correspond to the application to a transformed ODE system of one step of an implicit Runge-Kutta (IRK) method based on collocation with Gauss-Legendre nodes. As their underlying IRK methods, FCIRK methods are super-convergent: the method based on $s$ Gauss-Legendre nodes is of order $2s$.
Compared to symplectic splitting integrators, the implementation of FCIRK methods is more involved because an implicit system of equations has to be solved at each time-step. In return, such methods are better suited than symplectic splitting integrators 
(i) for exploiting parallel computer architectures (most of the computations can be done in $s$ processors in parallel),  (ii) for the mixed-precision approach mentioned above, as considerably fewer Keplerian motions in quadruple precision have to be computed for the same level of precision in a given integration interval, (iii) for the integration of equations including additional effects, such as post-Newtonian corrections, which cannot be written as the sum of integrable parts.

In our preliminary implementation of the 16th order FCIRK scheme~\cite{Anto2018},  time-steps of constant length were applied. However,  for long-term simulations of realistic models of the Solar System, the larger bodies of the asteroid belt (specially Ceres, Pallas, and Vesta) experience close encounters that have a significant impact in the chaotic behavior of the Solar System~\cite{Laskar2011b}. The precision of the numerical integration performed with any integrator with constant time-steps (in physical time) is degraded during close enough encounters, a degradation that may be more pronounced for the 16th order FCIRK method 
due to its higher order of precision and to the temporary loss of the hierarchy $H_I << H_K$ during close approaches.

Motivated by that,  we have analyzed  the effect of close encounters on the local discretization errors of our FCIRK integrators. More precisely, we have obtained rigorous estimates of the leading error term by applying results and techniques presented in~\cite{Anto2020}. 

Based on that analysis, we propose practical mechanism to identify integration steps that involve close encounters between arbitrary bodies. Actually, our analysis could be used as a theoretical basis to tray to endow our code with an adaptive time-stepping strategy.
However, it is known~\cite{Calvo1993} that the advantages of symplectic integrators for the long-term integration of Hamiltonian systems are lost if standard adaptive time-stepping strategies are used.

We adopt a different approach in FCIRK16. Our code is mainly intended for long-term integrations of the Solar System, where close encounters that seriously degrade the accuracy are rare.  In the absence of close approaches,  FCIRK16 advances in the integration with constant time-steps so that the integration enjoys the advantages of symplectic integration during long integration subintervals. In order to deal with eventual close encounters, FCIRK16  identifies and resolves in higher precision time-steps which would otherwise suffer from accuracy degradation due to the occurrence of a close approach.

This paper is organized as follows. Section~2 is devoted to define FCIRK methods and state their main properties. 
In Section~3, we analyze the effect of close encounters in the local discretization errors of FCIRK methods, and provide practical functions to monitor eventual close encounters. In Section~4, some aspects related to the implementation of FCIRK16 are briefly discussed. Section~5 is devoted to present some numerical experiments: In Subsection 5.1,  we check the results and the ideas of Section 3 with a 15-body point-mass Newtonian model of the Solar System (with the Sun, the eight planets, Pluto, and five main asteroids) and with a 16-body model where the Moon is treated as a separate body. In Subsection 5.2, we  present some numerical comparisons of FCIRK16 with a state-of-the-art high order explicit symplectic scheme that demonstrate the superiority of our integrator when very high precision is required.
We conclude with a summary of the work in Section~6.

\section{FCIRK methods for perturbed Kepler problems}

The equations of motion of the Hamiltonian function (\ref{eq:H}), written in terms of $q_i$, $v_i = p_i/\mu_i$,  read
\begin{equation}
\label{eq:qvODE}
\begin{split}
\frac{d}{dt} q_i &= v_i + g_i(u),\\
\frac{d}{dt} v_i &= -\frac{k_i}{\|q_i\|^3}q_i + g_{n+i}(u),
\end{split} \quad \mbox{for } i=1,\ldots,n,
\end{equation}
where $u=(q_1,\ldots,q_n,v_1,\ldots,v_n) \in \R^{6n}$,  and 
\begin{equation}
\label{eq:sw}
\begin{split}
 g_i(u) &=  \frac{\partial H_I}{\partial p_i}(q_1,\ldots,q_n,p_1,\ldots,p_n),\\
 g_{n+i}(u) &= -\frac{1}{\mu_i}\, \frac{\partial H_I}{\partial q_i}(q_1,\ldots,q_n,p_1,\ldots,p_n),
 \end{split}
\end{equation}
with $p_i$ ($i=1,\ldots,n$) replaced by $\mu_i\, v_i$ in the right-hand sides of (\ref{eq:sw}).

This can be more compactly written in the form
\begin{equation}
\label{eq:pertODE}
\frac{d}{dt} u = k(u) + g(u),
\end{equation}
where $k(u)$ corresponds to the Keplerian part and for each $u \in \R^{6n}$
\begin{equation*}
g(u) = (g_1(u),\ldots,g_{2n}(u)) \in \R^{6n}.
\end{equation*}
In what follows,  we will consider systems of the form (\ref{eq:pertODE}) with arbitrary maps $g_i(u)$ not necessarily of the form (\ref{eq:sw}) (so that (\ref{eq:pertODE}) may include dissipative terms).

We are concerned with the high-precision long-term numerical integration of  (\ref{eq:pertODE}) supplemented with the initial condition 
\begin{equation}
\label{eq:icond}
u(0) = u^0 := (q_1^0,\ldots,q_n^0,v_1^0,\ldots,v_n^0)  \in \R^{6n}.
\end{equation}

In FCIRK methods, as in the Wisdom and Holman (WH) integrator and its generalizations, one takes advantage of 
the integrability of the Keplerian equations of motion
\begin{equation}
\label{eq:Kepler_i}
\frac{d}{dt} q_i = v_i, \quad  \frac{d}{dt} v_i = -\frac{k_i}{\|q_i\|^3}q_i.
\end{equation}
That is,  one exploits the ability of efficiently computing for any $h \in \R$  (up to roundoff errors), the $h$-flow $\varphi_h:\R^{6n} \to \R^{6n}$ of the unperturbed system
\begin{equation}
\label{eq:keplerODE}
\frac{d}{dt} u = k(u).
\end{equation}
Recall that, by definition of $h$-flow,  $\varphi_{h}(u(t)) \equiv u(t +h)$ for  any solution $u(t)$ of (\ref{eq:keplerODE}).

Application of the time-dependent change of variables $u = \varphi_t(w)$ transforms  (\ref{eq:pertODE}) into the non-autonomous system 
\begin{equation*}
\frac{d}{dt} w = (\varphi'_t(w))^{-1}g(\varphi_t(w)).
\end{equation*}
From now on, we will use the notation
\begin{equation}
\label{eq:F}
F(w,t):= (\varphi'_t(w))^{-1}g(\varphi_t(w)).
\end{equation}
As shown in~\cite{Anto2018}, any solution $u(t)$ of (\ref{eq:pertODE}) satisfies the identity
\begin{equation*}
u(t+h) \equiv \varphi_{h/2}(\psi_h(\varphi_{h/2}(u(t)))),
\end{equation*}
where the map $\psi_h:\R^{6n} \to \R^{6n}$ is defined as follows: for arbitrary $u^0 \in \R^{6n}$, $\psi_h(u^0) = w(h)$, where $w(t)$ is the solution of the initial value problem
\begin{equation}
\label{eq:wODE}
\frac{d}{dt} w = F(w,t-h/2), \quad w(0)=u^0.
\end{equation}
Hence, $u(t+h)$ can be obtained from $u(t)$ by two Keplerian motions of time-increment $h/2$ alternated with one application of the map $\psi_h$ that accounts for the effect of the interactions.

In FCIRK methods, as in the WH map, the solution $u(t)$ of (\ref{eq:pertODE})--(\ref{eq:icond}) is approximated for $t= m\, h$, $m=1,2,3,\ldots$, by computing $u^m \approx u( m\, h)$ in a step-by-step manner as follows:  
for $m=1,2,3,\ldots$
\begin{equation}
\label{eq:FCIRK}
\begin{split}
w^m &= \varphi_{h/2}(u^{m-1}), \\ 
 \hat w^m &=\tilde \psi_h(w^m), \\
 u^m &= \varphi_{h/2}(\hat w^m),
\end{split}
\end{equation}
where the map $\tilde \psi_h$ is an approximation of the exact map $\psi_h$ accounting for the interactions.

In the WH integrator, $\tilde \psi_h(u^0)$ is defined for each $u^0 \in\R^{6n}$ as 
$\tilde \psi_h(u^0):= \tilde w(h)$, 
where $\tilde w(t)$ is the solution of the initial value problem
\begin{equation*}
\frac{d}{dt} \tilde w = g(\tilde w), \quad \tilde w(0)=u^0.
\end{equation*}
The WH map is a second order integrator because $\psi_h(u^0)  = \tilde w(h)  + \mathcal{O}(h^3)$ as $h \to 0$.

In a FCIRK method,  
$\tilde \psi_h(u^0)$
is defined as a Runge-Kutta approximation of  $\psi_h(u^0)=w(h)$ obtained
by applying one step of length $h$ to (\ref{eq:wODE}).  
More precisely, 
\begin{equation}
\label{eq:IRKPhi}
\tilde \psi_h(u^0) = u^0 + \Phi(u^0,h), \quad \Phi(u^0,h) = h \sum_{i=1}^s b_i \,  \dot W_i,
\end{equation}   
where the vectors $\dot W_{i}$ are implicitly defined  by
\begin{equation}
\label{eq:IRKdotW}
\dot W_{i} = F\left(u^0 + h \sum^s_{j=1} a_{ij}\, \dot W_j,  (c_j-1/2)\, h\right), \quad  i=1 ,\ldots, s.
\end{equation}
Here, $b_i$, $c_i$, $a_{ij}$ ($1\leq i, j \leq s$) are real parameters that determine the IRK scheme. The positive integer $s$ is referred to as the number of stages of the IRK method.
In~\cite{Anto2018}, we considered a particular family of time-symmetric symplectic IRK methods, the collocation methods based on $s$ Gauss-Legendre nodes.   More precisely, $c_1,\ldots,c_s$ are the zeros of $P_{s}(2x-1)$, where $P_s(x)$ is the Legendre polynomial of degree $s$, and  the coefficients $b_i$ ($1\leq i \leq s$) and $a_{ij}$ ($1\leq i,j \leq s$) are uniquely determined from the following equalities~\cite{HLW2006}:
 \begin{align}
\label{eq:B(s)}
\sum_{i=1}^s b_i c_i^{k-1} &= \frac{1}{k}, \quad k=1,\ldots,s, \\
\label{eq:C(s)}
\sum_{j=1}^s a_{ij}c_j^{k-1} &= \frac{c_i}{k}, \quad i=1,\ldots,s, \quad k=1,\ldots,s.
 \end{align}
Notice that  $c_1,\ldots,c_s$ and $b_1,\ldots,b_s$ are the nodes and weights respectively of the Gauss-Legendre quadrature formulas on the interval $[0,1]$.

The $s$-stage IRK method of collocation type with Gauss-Legendre nodes is of order $2s$. 
More precisely, 
\begin{equation}
\label{eq:local_error}
\tilde \psi_h(u^0) - \psi_h(u^0) = \frac{h^{2s+1}}{(2s+1)!}\, (\tilde w^{(2s+1)}(0)- w^{(2s+1)}(0)) +
\mathcal{O}(h^{2s+2}) 
\end{equation}
as $ h \to 0$, 
where $w^{(2s+1)}(0)$ denotes the $(2s+1)$th order derivative of the solution $w(t)$ of (\ref{eq:wODE}) evaluated at $t=0$, and $\tilde w^{(2s+1)}(0)$ denotes the $(2s+1)$th order derivative of the numerical solution $\tilde w(h):=\tilde \psi_h(u^0)$ evaluated at $h=0$.

In addition, if (\ref{eq:pertODE}) are 
 the equations of motion of a Hamiltonian function (\ref{eq:H}), then the map
 \begin{equation*}
 \tilde \psi_h:\R^{6n} \to \R^{6n}
 \end{equation*}
will be symplectic (as the map $\psi_h$ that is approximating to)
 (with respect to the 2-form $\sum_i \mu_i\, dq_i \wedge dv_i$), and hence also each step $u^{m-1} \mapsto u^m$ of the form (\ref{eq:FCIRK})
of the FCIRK method (see~\cite{Anto2018} for further geometric properties of FCIRK methods).


\section{Close encounters and local discretization errors}
\label{s:close_encounters}


We next try to estimate the effect of close encounters on the leading term of the local discretization error of 
 FCIRK methods.

\subsection{General analysis of local discretization errors}

  In view of (\ref{eq:local_error}), we will next aim at obtaining upper bounds of 
 \begin{equation}
 \label{eq:icondQV}
 \frac{1}{(2s+1)!} \|w_i^{(2s+1)}(0)\|, \quad i=1,\ldots,2n.
\end{equation}
(Here, we have collected the $(6n)$-vector $w(t)$ in $2n$ three-dimensional sub-vectors as $w(t) = (w_1(t),\ldots,w_{2n}(t))$.)
Such estimates should reflect how local discretization errors increase during close encounters.
  In order to do that, we will apply some results and techniques (based on the analytic extension to the complex domain of $N$-problems) presented in~\cite{Anto2020}.
 
Under the assumption that $g:\R^{6n} \to \R^{6n}$ is a real-analitic map, for each regular\footnote{In the sense that $g(u)$ is analytic in $u=u^0$ and that $|| q_i^0|| \neq 0$ for all $i$ (so that $k(u)$ is analytic in $u=u^0$).}
 initial state $u^0 \in \R^{6n}$, there exists $\rho=\rho(u^0)>0$ such that the following two conditions hold:
\begin{itemize}
\item[C1] The solution $u(t)=(q_1(t),\ldots,q_n(t),v_1(t),\ldots,v_n(t))$ 
  of (\ref{eq:pertODE})--(\ref{eq:icond}) and $g(u(t))$ can be uniquely extended as  analytic functions of $t$ in the closed disk 
\begin{equation*}
  \mathcal{D}_{\rho(u^0)}  = \{t \in \C\ : \ |t| \leq \rho(u^0)\}.
\end{equation*}
\item[C2]  For $i=1,\ldots,n$, 
\begin{equation*}
\|q_i(t) - q_i^0\| \leq \frac{1}{7}\, \|q_i^0\|.
\end{equation*}
\end{itemize}

In that case, it can be proven (we provide a proof in the Appendix) that, for $i=1,\ldots,n$,
\begin{align}
\label{eq:wi}
 \frac{1}{(2s)!} \|w_i^{(2s+1)}(0)\| &\leq  2\, \rho(u^0)^{-2s} (M_i(u^0) + \rho(u_0)\, M_{i+n}(u^0)) , \\
\label{eq:win}
 \frac{1}{(2s)!} \|w_{i+n}^{(2s+1)}(0)\| &\leq  \rho(u^0)^{-2s-1} (2\, M_i(u^0) + 3\, \rho(u^0)\, M_{i+n}(u^0)),
\end{align}
where for $i=1,\ldots,2n$,
\begin{equation}
\label{eq:Midef}
 M_i(u^0) = \sum_{t \in \mathcal{D}_{\rho(u^0)}} \|g_i(u(t))\|.
\end{equation}

\subsection{Local error estimates for planetary systems in canonical heliocentric coordinates}
\label{ss:error_heliocentric}

Next, we will obtain suitable $\rho(u^0)$ and $M_i(u^0)$ ($i=1,\ldots,2n$) as explicit functions of $u^0=(q_1^0,\ldots,q_n^0,v_1^0,\ldots,v_n^0)$ in one particular relevant case: the case where (\ref{eq:pertODE}) corresponds to a  point-mass $(n+1)$-body problem consisting of a massive body (the Sun) and $n$ bodies (major and minor planets, and asteroids) orbiting around it, written in Poincar\'e's canonical heliocentric coordinates.

Consider the Hamiltonian system corresponding to the $(n+1)$-body Hamiltonian function
\begin{equation}
  \label{eq:NbodyHam}
H(Q,P)=\frac{1}{2}\ \sum^{n}_{i=0}{\ \frac{{\|P_i\|}^2}{m_i}}-G\ \sum_{0\le i<j\le n}{\frac{m_im_j}{\|Q_i-Q_j\|}},
\end{equation}
written in cartesian coordinates $Q=(Q_0,\ldots,Q_n)$, $V=(V_0,\ldots,V_n)$, where $V_i=P_i/m_i$.
According to Theorem~2 in~\cite{Anto2020}  (applied with $\lambda=1/7$) the following statements S1--S2 hold for the Hamiltonian system supplemented with regular initial conditions 
\begin{equation}
\label{eq:icondQV}
Q_i(0) = Q_i^0, \quad V_i(0) = V_i^0, \quad i=0,1,\ldots,n.
\end{equation}
\begin{itemize}
\item[S1] It admits a unique analytic solution $(Q(t),V(t))$ defined in an open complex domain containing the closed disk $\{t \in \C\ : \ |t|\leq L(Q^0,V^0)^{-1}\}$, where 
\begin{equation}
\label{eq:L}
L(Q,V) =    \max_{0\leq i < j \leq n}  L_{ij}(Q,V)
\end{equation}
with
\begin{equation}
\label{eq:Lij}
  L_{ij}(Q,V) = \frac72\,\left(
\frac{\|V_i-V_j\|}{\|Q_i-Q_j\|} +  
\sqrt{\left( \frac{\|V_i-V_j\|}{\|Q_i-Q_j\|}\right)^2+
\frac{4}{7} \, \frac{K_{i}(Q) + K_j(Q)}{\|Q_i-Q_j\|}}\right),
\end{equation}
and
\begin{equation}
\label{eq:Ki}
K_i(Q) = \sum_{j\ne i,\ j=0}^{n}  \ \frac{G\, m_j}{\|Q_i-Q_j\|^2}, \quad i=0,1\ldots,n.
\end{equation}

\item[S2] In addition, if $|t| \leq L(Q^0,V^0)^{-1}$, then
\begin{equation*}
\|Q_i(t) - Q_j(t) - Q_i^0 +Q_j^0\| \leq \frac{1}{7}\, \|Q_i^0-Q_j^0\|, \quad 0 \leq i < j \leq n.
\end{equation*}
\end{itemize}

The Hamiltonian $H(q,p)$ obtained by rewriting the $(n+1)$-body Hamiltonian (\ref{eq:NbodyHam})  in terms of canonical heliocentric coordinates is of the form (\ref{eq:H}), with
  \begin{equation*}
H_I(q,p) := \frac{1}{m_0} \left(\sum\limits_{1 \leqslant i<j\leqslant n} p_i^T\ p_j \right)  -\sum\limits_{1 \leqslant i<j \leqslant n} \frac{G m_i m_j}{\|q_i-q_j\|},
\end{equation*}
and
\begin{equation*}
\frac{1}{\mu_i} = \frac{1}{m_i} + \frac{1}{m_0}, \quad 
k_i = G \, (m_0 +m_i), \quad 
i=1,\ldots,n. 
\end{equation*}
Here, we are assuming that the barycenter of the system is at rest, so that $q_0=(0,0,0)$ and $v_0=(0,0,0)$ can be ignored. The equations of motion, expressed in terms of the 3-vectors $q_i$, $v_i=p_i/\mu_i$, read (\ref{eq:qvODE}), with
 \begin{align}
 \label{eq:gi}
g_i(u) &= \sum_{j\ne i,\ j=1}^{n}  \frac{\epsilon_j}{1+\epsilon_j} v_j, \\
\label{eq:gni}
g_{i+n}(u) &=  -\sum_{j\ne i,\ j=1}^{n}  \ \frac{k_i\, \epsilon_j}{\|q_i-q_j\|^3} (q_i-q_j),
\end{align}
where
\begin{equation*}
\epsilon_i = \frac{m_i}{m_0}, \quad i=1,\ldots,n.
\end{equation*}
With that notation, the barycentric coordinates $(Q_i,V_i)$ ($i=0,1,\ldots,n$) are recovered from the canonical heliocentric coordinates $u=(q_1,\ldots,q_n,v_1,\ldots,v_n)$  as
\begin{align}
\label{eq:Heliocentric_to_Barycentric_0}
Q_0 &= -\sum_{i=1}^n \frac{m_i}{M}\, q_i, \quad V_0 = - \sum_{i=1}^n \frac{\epsilon_i}{1 + \epsilon_i} v_i, \\
\label{eq:Heliocentric_to_Barycentric_i}
Q_i &= Q_0 + q_i, \quad V_i = \frac{1}{1 + \epsilon_i} v_i, \quad i=1,\ldots,n.
\end{align}

The statements S1--S2 imply that C1--C2 hold for 
\begin{equation}
\label{eq:rho}
\rho(u) =   \min_{0\leq i < j \leq n} \rho_{i,j}(u)
\end{equation}
with 
\begin{equation}
\label{eq:rhoij}
\rho_{i,j}(u) = L_{i,j}(Q,V)^{-1}
\end{equation}
where (\ref{eq:Heliocentric_to_Barycentric_0})--(\ref{eq:Heliocentric_to_Barycentric_i}) is used to compute  the barycentric coordinates $(Q,V)$ from $u$.
We next give upper bounds for (\ref{eq:Midef}) required to get the estimates (\ref{eq:wi})--(\ref{eq:win}) of the $(2s+1)$th derivatives $w^{(2n+1)}(0)$ featuring in the leading term of the local discretization error (\ref{eq:local_error}).

The statements S1--S2 imply that, for each $t \in \mathcal{D}_{\rho(u^0)}$,
\begin{align}
\label{eq:qij}
\|q_i(t)-q_j(t) - q_i^0+ q_j^0\| &\leq \frac{1}{7}\, \|q_i^0-q_j^0\|, \quad 1 \leq i < j \leq n, \\
\label{eq:vit}
\|v_i(t)-v_i^0\| &\leq 2\, (1 + \epsilon_i) \, \rho(u^0) \, K_i(u^0), \quad  1 \leq i \leq n.
\end{align}
From (\ref{eq:vit}), one gets
\begin{equation}
\label{eq:Mi}
M_i(u^0) \leq \sum_{j\ne i, \ j=1}^n \frac{\epsilon_j}{1+\epsilon_j} \|v_j^0\| + 2\, \rho(u^0) \sum_{j\ne i, \ j=1}^n\epsilon_j \, K_j(u^0), \quad  1 \leq i \leq n.
\end{equation}
 Lemma~1 in~\cite{Anto2020} and (\ref{eq:qij}) finally imply that
\begin{equation}
\label{eq:Min}
M_{i+n}(u^0) \leq  2\, \sum_{j\ne i,\ j=1}^{n}  \ \frac{k_i\, \epsilon_j}{\|q_i^0-q_j^0\|^2}, \quad  1 \leq i \leq n.
\end{equation}

 \subsection{The effect of close encounters in the local discretization errors}
\label{ss:effect_close_encounters}

Consider, as in previous Subsection, an $(n+1)$-body system described in canonical heliocentric coordinates, and assume that the mass ratios $\epsilon_i=m_i/m_0$ ($i=1,\ldots,n$) are small.

We want to study the effect of close encounters on the local discretization errors $\delta(w^m,h) = \tilde \psi_h(w^m)-\psi_h(w^m)$, $m=0,1,2,3,\ldots$ made during the FCIRK integration (\ref{eq:FCIRK}). For each $i=1,\ldots,n$, we denote as $\delta_{i}(u^0,h)$ (resp. $\delta_{i+n}(u^0,h)$)  the 3-vector that collects the components of $\delta(u^0,h)$ corresponding to the positions (resp. velocities) of the $i$th body.

   While integrating with FCIRK, whenever all the bodies orbiting around the central massive body are well separated in $w^m$ (we next set $u^0:=w^m$),
\begin{itemize}
\item  the upper bounds (\ref{eq:Mi})--(\ref{eq:Min}) will remain small, and 
\item typically, $\rho(u^0)=\rho_{0,1}(u^0)$, where $i=0$ corresponds to the central massive body (say, the Sun) and body $i=1$ is nearest the central body (say, Mercury). 
\end{itemize}
In that case, $\rho(u^0)$ will oscillate with approximately the orbital period of Mercury, with lowest values of $\rho(u^0)$ at Mercury's perihelion. In view of (\ref{eq:local_error}) and 
 (\ref{eq:wi})--(\ref{eq:win}),  the components of the local discretization errors are expected to oscillate likewise. 
 
 However, if  body $i=k$ and $j=\ell$ are close enough to each other in $u^0$, then $\rho_{k,\ell}(u^0)$ will be smaller than the value of $\rho_{0,1}(u^0)$ at Mercury's perihelion, and thus
  $\rho(u^0)=\rho_{k,\ell}(u^0)$. (In addition, among the upper bounds  (\ref{eq:Mi})--(\ref{eq:Min}), those for $i=k$ and $i=\ell$ will dominate over the rest for close enough encounters between bodies $k$ and $\ell$. )
  
 We thus expect that the local error
 will grow roughly as $\sim (h/\rho(w^m))^{2s+1}$ when $\rho(w^m)$ decreases due to a close encounter.  
  This suggest that 
  \begin{itemize}
\item   close encounters that may degrade the accuracy of the FCIRK integration (\ref{eq:FCIRK}) can be identified by monitoring the value of $\rho(w^m)$, and
\item in such cases, the local discretization could be controlled by decreasing the step size $h$ of the IRK scheme as $\rho(w^m)$ decreases to keep $h/\rho(w^m)$ roughly constant.
   \end{itemize}
    
\subsection{The case of models of the Solar System with the Moon as a separate body}
\label{ss:error_with_Moon} 
 
 The discussion in previous two subsections was applicable to models of the Solar System where the Earth-Moon system is treated as a point mass located at their barycenter. 
 We now focus on the case of $(n+1)$-body models of the Solar System with the Earth and the Moon treated as two separate point masses, say, bodies $i=n-1$ and $i=n$ respectively.


We consider canonical Heliocentric coordinates for bodies $i=1,2,\ldots,n-2$ and for the Earth-Moon barycenter, and  geocentric positions and corresponding conjugate momenta for the Moon. The $(n+1)$-body problem with Hamiltonian (\ref{eq:NbodyHam}) described in such canonical coordinates is of the form (\ref{eq:H}), where
\begin{align*}
\frac{1}{\mu_i}&=\frac{1}{m_0}+\frac{1}{m_i}, \quad k_i=G\, (m_0+m_i), \quad i=1,\ldots,n-2,  \\
\frac{1}{\mu_{n-1}}&=\frac{1}{m_0}+\frac{1}{(m_{n-1}+m_{n})}, \quad 
k_{n-1}=G\, (m_0+m_{n-1}+m_{n}), \\
\frac{1}{\mu_{n}}&=\frac{1}{m_{n}}-\frac{1}{(m_{n-1}+m_{n})}, \quad
k_{n}=\frac{G\, m_{n-1}^3}{(m_{n-1}+m_{n})^2}.
\end{align*}
and
\begin{align*}
H_I(q,p)=&\frac{1}{m_0}\sum_{1 \leqslant i < j}^{n-1} p_i^T\,  p_j - \sum_{1 \leqslant i < j}^{n-2}\frac{G\, m_i\, m_j}{\|q_i-q_j\|}\\
&-\sum_{i=1}^{n-2}\left(\frac{G\, m_i\, m_{n}}{\|q_i-q_{n-1}-q_{n}\|}+ \frac{G\, m_i\, m_{n-1}}{\|q_i-q_{n-1}+\frac{G\, m_{n}}{m_{n-1}}q_{n}\|}\right) \\
&+\frac{G\, m_0 (m_{n-1}+m_{n})}{\|q_{n-1}\|}-\frac{G\, m_0\, m_{n-1}}{\|q_{n-1}-\frac{G\, m_{n}}{m_{n-1}}q_{n}\|}-\frac{G\, m_0\, m_{n}}{\|q_{n-1}+q_{n}\|}.
\end{align*}

The barycentric coordinates $Q_i$, $V_i = P_i/m_i$, $i=0,1,\ldots,n$, are recovered from the coordinates $q_i$, $v_i=p_i/\mu_i$, $i=1,\ldots,n$, as follows:
\begin{equation}
\label{eq:qv2QV}
\begin{split}
Q_i &= Q_0 + q_i, \quad V_i = \frac{\mu_i}{m_i} v_i, \quad i=1,\ldots,n-2, \\
Q_{n-1} &= Q_0 + q_{n-1}-\frac{m_{n}}{m_{n-1}}q_{n}, \quad Q_n = Q_0 + q_{n-1} + q_n, \\
V_{n-1} &= \frac{\mu_{n-1}}{m_{n-1} + m_n} \, v_{n-1} - \frac{\mu_n}{m_{n-1} + m_n}\,  v_n, \\
V_n &=  \frac{\mu_{n-1}}{m_{n-1} + m_n} \, v_{n-1} + \frac{m_{n-1} \mu_n}{m_n\, (m_{n-1} + m_n)} \, v_n, \\ 
Q_0 &= -\sum_{i=1}^{n-2} \frac{m_i}{M}\, q_i - \frac{m_{n-1}+m_n}{M} \,q_{n-1}, \quad 
V_0 = - \sum_{i=1}^{n-1} \frac{\mu_i}{m_0} v_i.
\end{split}
\end{equation}

It is not difficult to prove, as in Subsection~\ref{ss:error_heliocentric}, that C1--C2 also hold in that case,  with $\rho(u)$ given by (\ref{eq:rho})--(\ref{eq:rhoij}), where $L_{i,j}(Q,V)$ is given by (\ref{eq:Lij}) and (\ref{eq:qv2QV}) is applied to compute  the barycentric coordinates $(Q,V)$ from $u$.   Upper bounds of $M_i(u)$ similar to those obtained in Subsection~\ref{ss:error_heliocentric} can also be obtained. The local error (\ref{eq:local_error}) can thus be estimated with the help of (\ref{eq:Mi})--(\ref{eq:Min}). 

We eventually arrive at the conclusion that the local error
 will grow roughly as $\sim (h/\rho(w^m))^{2s+1}$ when $\rho(w^m)$ decreases due to some close encounter. During most of the simulation, $\rho(w^m) = \rho_{n-1,n}(w^m)$ due to the highly oscillatory nature of the orbit of the Moon, and only for very extreme close encounters  between body $k$ and $\ell$ does $\rho(w^m)$ decrease noticeably (as $\rho_{k,\ell}(w^m)$ becomes smaller than $\rho_{n-1,n}(w^m)$ and consequently $\rho(w^m)<\rho_{n-1,n}(w^m)$).
 However,  we have observed in practice that  some close encounters that degrade the accuracy of the simulation are not close enough to have $\rho(w^m)<\rho_{n-1,n}(w^m)$, and hence cannot be detected by monitoring the value of $\rho(w^m)$. We believe that this is due to the fact that the contribution of the high oscillation of the Moon on the local discretization errors of the rest of the bodies is typically small. Moreover,  the contribution of such highly oscillatory local errors over accumulated errors are further reduced because they are partially averaged out. 
 
 Motivated by that, we propose to modify the function $\rho(u)$ for monitoring close encounters when the Moon is treated as a separated body by excluding $\rho_{n-1,n}(u)$ from the definition of $\rho(u)$.

 \section{Implementation aspects}
\subsection{Implementation of the IRK approximation}
\label{ss:implementationIRK}

In our implementation of FCIRK methods, we compute the vectors $\dot W_i$ from (\ref{eq:IRKdotW}) by using the fixed-point implementation of IRK schemes presented in~\cite{Anto2017}. This consists of a robust and efficient implementation that avoids the systematic accumulation of errors in energy (due to round-off errors and iteration errors) observed in standard fixed-point implementations of symplectic implicit RK schemes~\cite{Hairer2006,Anto2017}. 

We compute the initial guesses to start  the fixed point iteration with an extrapolation procedure that allows us to substantially reduce the required number of fixed point iterations for small enough time-steps $h$. However, for FCIRK methods with large number $s$ of implicit stages, this is only true for extremely small values of the step-size. Furthermore, the extrapolation formulae become numerically  unstable (due to cancellation errors caused by large coefficients of opposite sign) for more large values of $s$.  These two factors limit, for practical computations in 64-bit or 80-bit floating point arithmetic, the efficiency of FCIRK methods for large values of $s$, say $s\geq 12$. After some preliminary numerical experiments, we have concluded that $s=6,7,8,9$ are good choices in this respect. Similar considerations apply to implicit Runge-Kutta schemes of collocation type: in~\cite{HLW2006}, the choice $s=6$ is favored for 64-bit floating point arithmetic in the case  of Gauss-Legendre nodes, and $s=8$ is chosen in~\cite{Everhart1985,Rein2014} in the case of Gauss-Radau nodes. For FCIRK methods, we finally choose the scheme with $s=8$ stages (hence of order $2s=16$) because it allows a more efficient parallelization with four or eight threads.   

We refer to our implementation of the $16$th-order FCIRK method as FCIRK16. 
In FCIRK16, the evaluation of the right-hand side of (\ref{eq:IRKdotW}) for each $i \in \{1,\ldots,8\}$  is optionally performed in parallel (with openMP) with a prescribed number of threads.
\subsection{Mixed precision implementation of FCIRK16}
\label{ss:mixedprecision}

A mixed precision approach proposed in~\cite{Anto2018} to reduce the effect of using finite precision arithmetic is implemented in FCIRK16.  The main idea consists of applying most of the computations in a basic  floating point arithmetic, and perform a few critical operations in a higher precision arithmetic.

In our current implementation of FCIRK16, $ \hat w^m =\tilde \psi_h(w^m)$ in (\ref{eq:IRKPhi}) is evaluated as 
\begin{align}
\nonumber
\Delta^m &= \Phi(w^m,h), \\
\label{eq:sumwm}
\hat w^m &= w^m + \Delta^m,
\end{align}
where $\Phi(u^m,h)$ is evaluated in $80$-bit extended precision, while the rest of the computations in (\ref{eq:FCIRK}), that is, the evaluations of the map $\varphi_{h/2}$ and the sum (\ref{eq:sumwm})
are performed in quadruple precision ($128$-bit floating point arithmetic).
This is motivated by the following:
\begin{itemize}
 \item The evaluation of $\Phi(w^m,h)$ represents the bulk of the computations of each step (\ref{eq:FCIRK}). Note that this is true regardless of the complexity of the evaluation of the interaction term $g(u)$ in (\ref{eq:pertODE}). Indeed, if $m$ fixed point iterations are needed at each step (typically, $m=5,6$), $sm=8m$ evaluations of $\varphi_h$ and its partial derivatives (in addition to $8m$ evaluations of $g(u)$) are required to evaluate $\Phi(w^m,h)$, compared to two evaluations of $\varphi_h$ (and $6n$ additions) in quadruple precision. Hence, the increase in computing time of one step  (\ref{eq:FCIRK})  due to the mixed precision implementation compared to one step fully implemented in $80$-bit extended precision is always relatively small.
  \item The components of the vector $\Delta^m$ in (\ref{eq:sumwm}) are typically of considerably smaller magnitude than the corresponding component of $w^m$.  If their magnitude is, say, $2^{-k}$ smaller, then the overall precision of each step (\ref{eq:FCIRK}) will correspond to a significand of $64+k$ bits compared to a precision of $64$-bit significand one would have if all the computations in (\ref{eq:sumwm}) were performed in $80$-bit floating point arithmetic.
\end{itemize}

\subsection{Dealing with close encounters}
\label{ss:adaptivity}

  We aim at enjoying the favorable behavior of constant time-step symplectic  integration  during long integration subintervals, and only occasionally (during eventual close encounters) perform a more dedicated (computationally more expensive) numerical integration to compute $u^m \approx u( m h)$ from $u^{m-1}$.
We thus have endowed FCIRK16 with a mechanism to detect the time-steps (from now on, {\em critical time-steps}) involving a close encounter that would substantially   degrade the accuracy of the numerical integration in constant-step size mode.

In such critical time-steps, $ \tilde \psi_h(w^m)$ in (\ref{eq:FCIRK}) is replaced by a better approximation of $\psi_h(w^m)$. While in ordinary steps $ \hat u^m = \tilde \psi_h(w^m)$ is obtained by applying one step of length $h$ of the underlying IRK scheme applied to the transformed system, in critical steps, $\hat u^m$ will be computed by applying $k$ steps of constant length $h/k$ of the IRK scheme to the transformed system. 
In addition, since the mixed precision approach becomes less effective during close encounters (as the dominance of the Keplerian Hamiltonian $H_K$ over the interaction Hamiltonian is reduced or lost during such events), we implement critical time-steps fully in quadruple precision $128$-bit floating point arithmetic.

We next specify the criterion to identify critical time-steps that we have implemented in FCIRK16, and the rule used to choose the number $k$ of substeps of the IRK schemes to be applied in each critical step.  Both tasks are done with the help of the monitoring function $\rho(u)$ defined in Subsection~\ref{ss:error_heliocentric} for $(n+1)$-body problems described in heliocentric coordinates, and the modified monitoring function (that we will also refer to as $\rho(u)$) defined at the end of Subsection~\ref{ss:error_with_Moon} for the case where the Moon is treated as a separate body.  For systems of the form (\ref{eq:pertODE}) obtained by describing an $(n+1)$-body problem in other coordinates the monitoring function $\rho(u)$ could be defined similarly.

In FCIRK16, the $m$th step (\ref{eq:FCIRK}) will be considered critical if, once $w^m=\varphi_{h/2}(u^{m-1})$ is computed
 \begin{equation}
 \label{eq:criterion_critical_steps}
\rho(w^m) <  \mu - \nu \, \sigma,
\end{equation}
where $\mu$ (resp. $\sigma$) is the arithmetic mean  (resp. the standard deviation) of the values of $\rho(w^k)$ for all previous ordinary steps, and  $\nu$ is a prescribed positive number. We set $\nu=1.6$ as default value in FCIRK16.

Motivated by second item at the end of Subsection~\ref{ss:effect_close_encounters}, we choose the number $k$ of substeps of constant length $h/k$ of the IRK scheme to be performed when computing $\hat w^m$ from $w^m$,  as
the positive integer $k$ such that
 \begin{equation}
 \label{eq:kcond}
   k-1 < \frac{\mu}{\rho(w^m)} \leq k.
 \end{equation}

 \section{Numerical experiments}
\label{s:NumericalExperiments}

We next present some numerical experiments: The experiments in the first subsection are aimed to assess the effectiveness of the technique for monitoring close encounters introduced in Section~\ref{s:close_encounters}. In the second subsection,  we demonstrate the performance of FCIRK16 for high precision long-term integrations of the Solar System by comparing it with a state-of-the-art explicit symplectic integrator.

We consider two Newtonian point-mass $(n+1)$-body models of  the Solar System: From one hand, a 15-problem with the Sun, the eight planets, Pluto, 
and five main bodies of the asteroid belt (Ceres, Pallas, Vesta, Iris and Bamberga) described in Poincare's canonical heliocentric coordinates.  On the other hand, we consider the 16-problem obtained from the former by considering the Moon as a separate body, with canonical coordinates described in Subsection~\ref{ss:error_with_Moon}.

We consider the initial values at Julian time (TDB) 2440400.5 (the 28th of June of 1969), from the DE430 Ephemerides \cite{Folkner2014}. 

All the numerical tests presented here are run in a HP-Z6-G4-Workstation with 8 cores of  Intel\copyright Xeon\copyright Silver 4110 CPU @ 2.10GHz and 32GiB (16 +16 DIMM DDR4) of RAM memory. The code is written in C and has been compiled under  Ubuntu 18.04.3 LTS operating system with gcc  version 7.5.0 with options -O2 -std=c99 -fno-common -mfma. The parallelized implementation of FCIRK16 is based on the OpenMP library. 

\subsection{Monitoring close encounters}

We first conduct some experiments to check the techniques of monitoring close encounters proposed and discussed in Section~\ref{s:close_encounters}. 
For that purpose, we have first  made two integrations of the 15-body problem with FCIRK16 forward in time for $50000$ years with a constant time-step of $h=1.5$ days and $h=3$ days respectively. 
\begin{figure}[ht!]
\centering
{\includegraphics[width=0.95\textwidth]{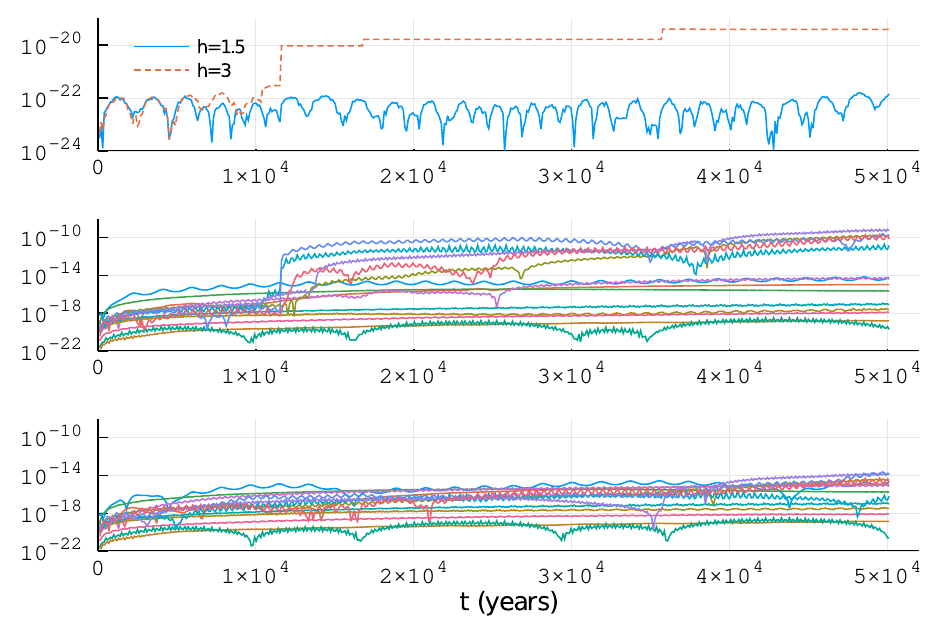}} 
\caption{\small
Evolution of relative errors in energy for FCIRK16 with $h=1.5$ and $h=3$ (top), and evolution of errors in position of each of the body for the non-adaptive (resp. adaptive with $\nu=1.6$) implementation of FCIRK16 with $h=1.5$ in the middle (resp. lower) subplot.
\label{fig:1}}
\end{figure}

 The evolution of relative errors in energy is displayed in the upper subplot of Figure~\ref{fig:1}, while the middle (resp. lower) subplot shows the evolution of the errors in position \footnote{We compute the actual position errors by considering as exact the results obtained with constant step-size $h=0.5$ with a version of FCIRK16 fully implemented in quadruple precision.} for each of the bodies orbiting around the Sun for the non-adaptive  (adaptive) integration with constant step-size $h=1.5$. 
  
Some jumps in the energy errors are clearly observed in the integration with $h=3$. (The largest jump coincides with a close encounter between Pallas and Vesta.) On the contrary,  there is no clear sign of accuracy degradation in the plot of the evolution of relative errors in energy for $h=1.5$. However, a substantial loss of precision in the positions of some of the bodies can clearly be seen in the middle subplot of Figure~\ref{fig:1}. We have also made an analogous experiment for the 16-body problem, and obtained very similar plots. We conclude that monitoring the energy error is  not enough to identify critical close encounters between minor bodies in the asteroid belt. This is due to the marginal contribution of the asteroids to the total energy of the system.  

The lower subplot of Figure~\ref{fig:1} shows that the technique to reduce the accuracy loss due to close encounters proposed in Subsection~\ref{ss:adaptivity} has been effective in this example. We have used the default value $\nu=1.6$, which has resulted in 203 critical steps out of $122\times 10^5$ ordinary steps. It is worth stressing that,  due to the low percentage of critical steps, applying our adaptive technique does not have any practical impact in the execution time of the integration.

\begin{figure}[ht!]
\centering
{\includegraphics[width=0.95\textwidth]{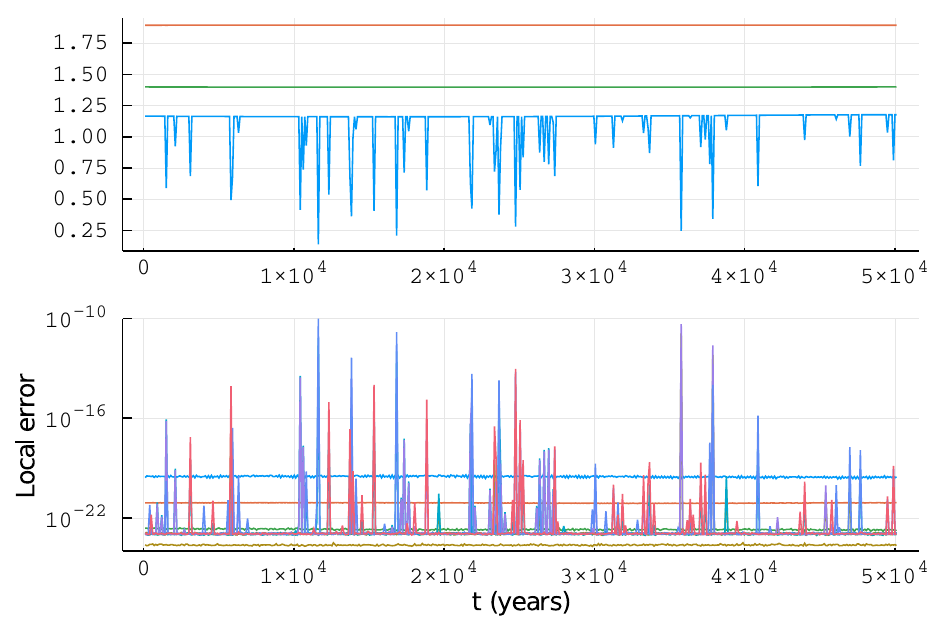}}
\caption{\small
Close encounter monitoring for the 15-body problem: (a) Curves for mean value $\mu$ of $\rho$, $\mu-\sigma$ (where $\sigma$ is the standard deviation of $\rho$), and $\min \rho$  (top), (b) evolution of local errors for each of the orbiting bodies (bottom)
\label{fig:2}}
\end{figure}

We next take a closer look to  the effectiveness of the monitoring technique proposed in Section~\ref{s:close_encounters}. Figure~\ref{fig:2} corresponds to an integration of 
 the 15-body problem forward in time for $50000$ years with a constant time-step of $h=6$ days. 
 (We have saved the results of the integration every $6100$ steps, that is, every 36600 days.)  
\begin{itemize}
\item The upper  subplot displays the evolution of $\rho(w^m)$. Three curves are drawn in that subplot: The mean value $\mu$  (since the beginning of the integration interval) of $\rho(w^m)$, the difference $\mu-\sigma$, where $\sigma$ is standard deviation of $\rho(w^m)$, and the minimum of $\rho(w^m)$ over subintervals of 36600 days. Several sharp minima are observed in the later cuve, corresponding to encounters between different asteroids.  

\item The lower subplot displays the evolution of estimated local errors in positions of each of the bodies (they are computed as the difference of $u^m$ with the more precise approximation obtained by applying two steps of length $h/2$ of the FCIRK scheme starting from $u^{m-1}$). 
\end{itemize}
The sharpest spikes of local error (corresponding to close encounters between asteroids) nicely coincide in time with the sharpest local minima of the monitoring function.  We conclude that the proposed monitoring technique works very well in this case.

\begin{figure}[ht!]
\centering
{\includegraphics[width=0.95\textwidth]{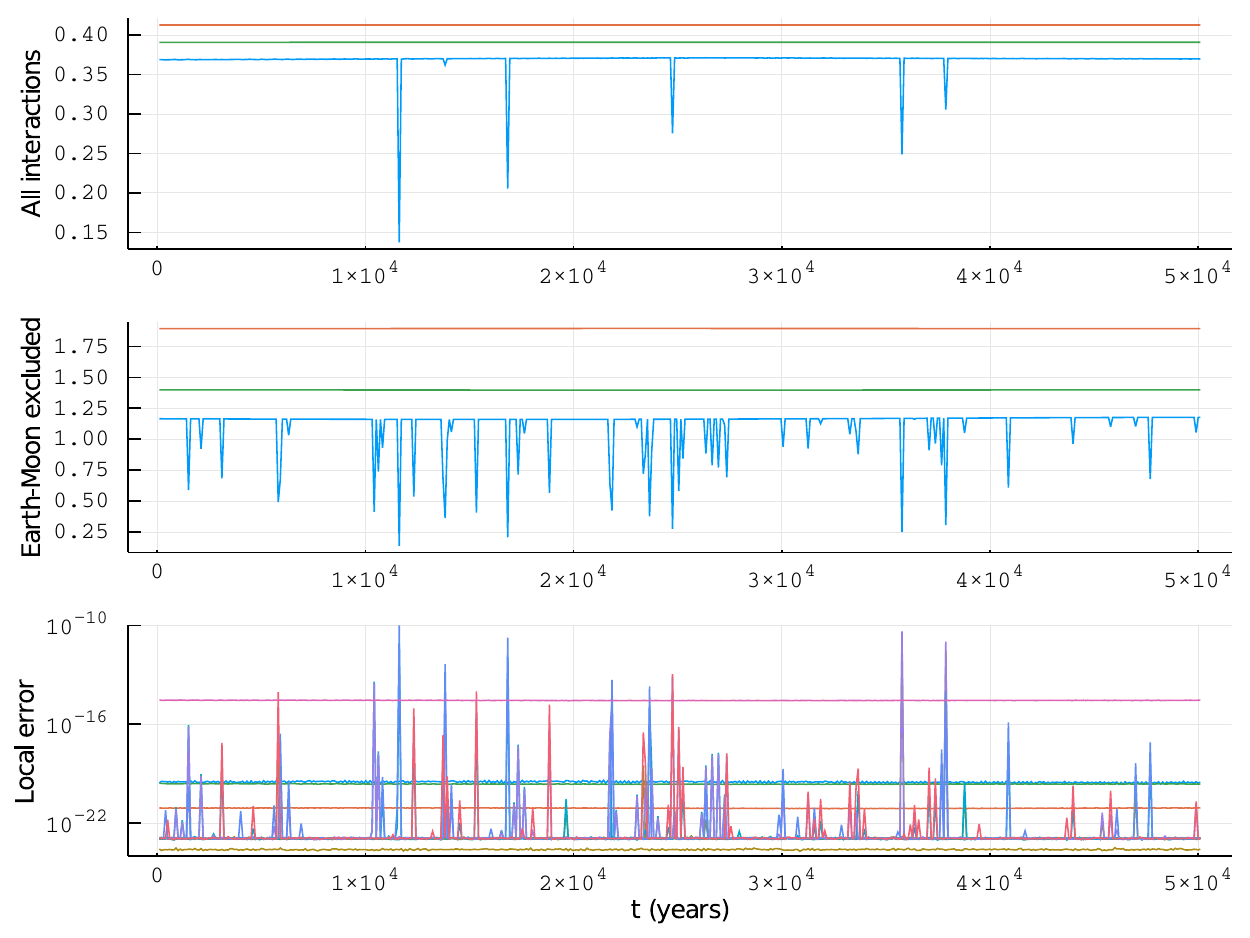}}
\caption{\small
Close-encounter monitoring for the 16-body problem: (a) Curves for mean value $\mu$ of $\rho$, $\mu-\sigma$ (where $\sigma$ is the standard deviation of $\rho$), and $\min \rho$  with all interactions (top), (b) Curves for  $\mu$, $\mu-\sigma$, and $\min \rho$ with the Earth-Moon interaction excluded (middle), (c) 
evolution of local errors for each of the orbiting bodies (bottom)
\label{fig:3}}
\end{figure}

In Figure~\ref{fig:3}, the results of a similar experiment performed for  the 16-body problem are reported. 
This time, the three curves ($\mu$, $\mu-\sigma$, and $\min \rho$) corresponding to  two different monitoring functions are displayed: (i) the original monitoring function in the upper subplot, (ii) and the modified monitoring function proposed at last item at the end of Subsection~\ref{ss:error_with_Moon} (which excludes the Earth-Moon interaction)  in the middle subplot. In the lower subplot, the evolution of the estimated local errors in positions of each of the bodies are displayed. The upper nearly horizontal curve corresponds to the local error in position of the Moon. (The local errors in position of Mercury and the Earth are five orders of magnitude smaller.)  The sharp local minima of the original monitoring function identify correctly the spikes of local error  that arise well above the local error of the Moon, but some of the less pronounced spikes are not reflected in the original monitoring function. The modified monitoring function, displayed in the middle subplot, is able to identify nicely the spikes of local error that arise above Mercury's local error.

\subsection{Comparison with a state-of-the art explicit symplectic integrator}

The efficiency of several generalizations of the classical Wisdom and Holman integrators (in Jacobi coordinates) for high precision long-term integration of the Solar System are compared in~\cite{Rein2019}. It is there concluded that, for high precision integrations, the symplectic integrator ABA(10,6,4)   presented in \cite{Blanes2013,Farres2013} is the most efficient among all the considered methods.

The integrator ABA(10,6,4) requires that the interaction Hamiltonian $H_I$ only depend on the positions (or at least, that $H_I$ can be written as the sum of two commuting integrable Hamiltonians). This is for instance the case if Jacobi coordinates are used to rewrite the $N$-body Hamiltonian (\ref{eq:NbodyHam}) in the form 
(\ref{eq:H}). If Poincare's canonical Heliocentric coordinates are used instead,  the interaction Hamiltonian is of the form 
\begin{equation}
\label{eq:HIAB}
H_I(p,q) = A(q) + B(p). 
\end{equation}
 In \cite{Blanes2013}, the integrator ABAH(10,6,4) is constructed as an alternative to ABA(10,6,4) when Poncar\'e's Heliocentric coordinates are used instead of Jacobi coordinates (more generally, for interaction Hamiltonians of the form (\ref{eq:HIAB})).  In the numerical integrations of the Solar System performed in~\cite{Farres2013}, it is concluded that ABAH(10,6,4) applied with Heliocentric coordinates is as efficient as  ABA(10,6,4) applied with Jacobi coordinates.

 The three positive integers in the name of the integrators  ABA(10,6,4) and ABAH(10,6,4) refer to their global error estimates, which are in both cases of the form $\mathcal{O}(\epsilon\, h^{10} + \epsilon^2\, h^6 + \epsilon^3\, h^4)$, where $\epsilon$ is the maximum of the mass ratios $m_i/m_0$. In contrast, the global error of FCIRK16 is of the form $\mathcal{O}(\epsilon\, h^{16})$.
 
In the case of the two models of the Solar system that we are considering in our numerical experiments (that is,  the interaction Hamiltonians described in Subsection~\ref{ss:error_heliocentric} and Subsection~\ref{ss:error_with_Moon},) are also of the form (\ref{eq:HIAB}). We will thus compare FCIRK16  with the symplectic integrator ABAH(10,6,4) presented in \cite{Blanes2013,Farres2013}.

The splitting symplectic integrator ABAH(10,6,4) is implemented by alternating evaluations of the Keplerian motions with evaluations of the gradients of $A(q)$ and $B(p)$ in (\ref{eq:HIAB}). No symplectic correction is applied.
 We apply our own optimized implementation of the later in $80$-bit extended precision arithmetic (with  Kahan's compensated summation~\cite{Kahan1965}), which makes use of the same routines for the evaluation of $\varphi_h$ and the gradient of the interaction Hamiltonian as in  FCIRK16.    We do not attempt the implementation of parallelization techniques for ABAH(10,6,4):  splitting integrators, unlike the FCIRK methods, require the sequential evaluation of the gradients of the interaction Hamiltonians. Consequently, parallelization  of ABAH(10,6,4) is only possible inside the evaluation of the gradient, hence with a lower granularity than that of FCIRK16, with would have a considerably  higher parallelization overhead.

 The technique proposed in Subsection~\ref{ss:adaptivity} can easily be adapted to splitting symplectic integrators and in particular to the method ABAH(10,6,4).  For a fair comparison with FCIRK16, we have implemented ABAH(10,6,4) with a similar mechanism to deal with close encounters.  In order to illustrate the advantage of incorporating that technique into our implementation of ABAH(10,6,4), we have made two integrations of the 15-body problem considered in the previous subsection with step-size $h=1$, one with fully constant step-size, and the other one with our technique for detecting and coping with close encounters. In the upper plot of Figure~4, we display the evolution of the position errors of each of the orbiting bodies for the integration with fully constant step-size $h=1$. One can observe a clear degradation of the precision of the positions of some of the bodies due to close encounters.
In the lower plot of Figure~4, we can see the evolution of the errors in the integrations corresponding to $\nu=1.6$ (304 critical steps out of $183\times 10^5$ ordinary steps have been required in that case).

\begin{figure}[ht!]
\centering
{\includegraphics[width=.95\textwidth]{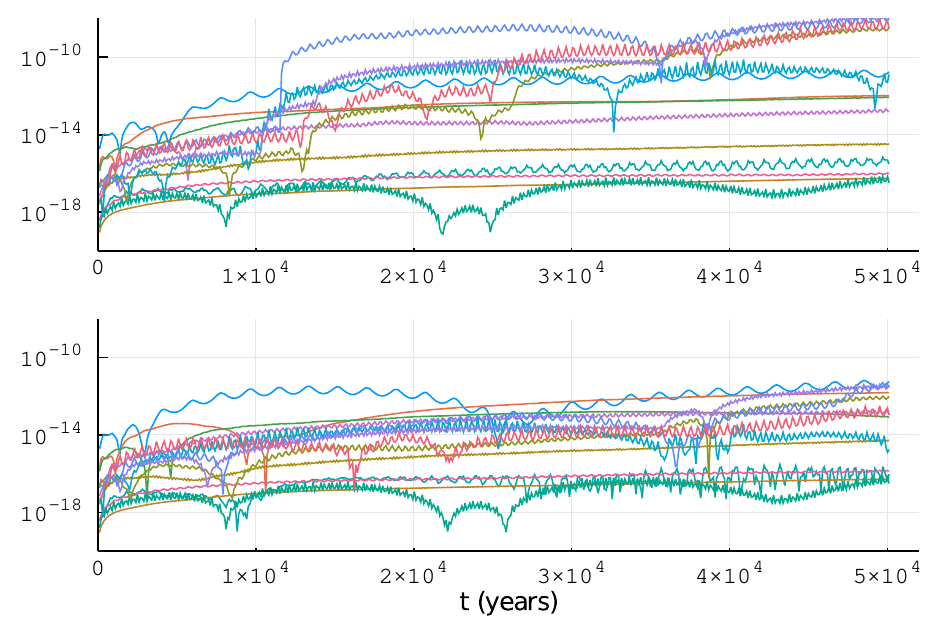}} 
\caption{\small
Evolution of position errors for ABAH1064 applied with $h=1$ to the 15-body problem: fully constant time-step (top), adaptive implementation with $\nu=1.6$ (bottom)
\label{fig:4}}
\end{figure}

In~\cite{Anto2018}, the efficiency of a preliminary implementation of our 16th order FCIRK method was compared to that of ABAH(10,6,4) (and also to the 16th order implicit Runge Kutta implemented in~\cite{Anto2017}). A simple 10-body model of the Solar System (including no asteroids, nor the Moon as a separate body) was considered.  In particular, efficiency diagrams displaying the error in energy versus execution time for different values of the constant time-step were presented for that purpose. We have repeated such experiments for the 15-body model and 16-body model of the Solar System described above for our current implementation of FCIRK16 (with the default value $\nu=1.6$ in (\ref{eq:criterion_critical_steps})). We have obtained very similar efficiency diagrams (not shown here), the main difference (compared to those presented in~\cite{Anto2017}) being that, due to the higher complexity of the evaluation of the gradient of the interaction Hamiltonian, higher parallelization speedup is obtained for the execution of FCIRK16. The parallel execution  of FCIRK16 with four threads is (both in the 15-body problem and the 16-body problem) approximately twice as fast as the sequential execution.

As in~\cite{Anto2018}, we conclude that ABAH(10,6,4) is more efficient than FCIRK16 for lower precisions.
However, if higher precision is required, ABAH(10,6,4) is limited by the precision of the 80-bit floating point arithmetic. At that limit, thanks to its mixed precision implementation, FCIRK16 is able to reduce errors by two orders of magnitude with similar computing time.  We have also tested a mixed-precision implementation ABAH(10,6,4) that computes all the Keplerian flows in quadruple precision, but its execution time increases by a factor between 6 and 9 in our examples. This is an expensive toll to be paid for a little of extra precision, while in FCIRK16, this is achieved  essentially for free. 

\begin{figure}[ht!]
\centering
{\includegraphics[width=.95\textwidth]{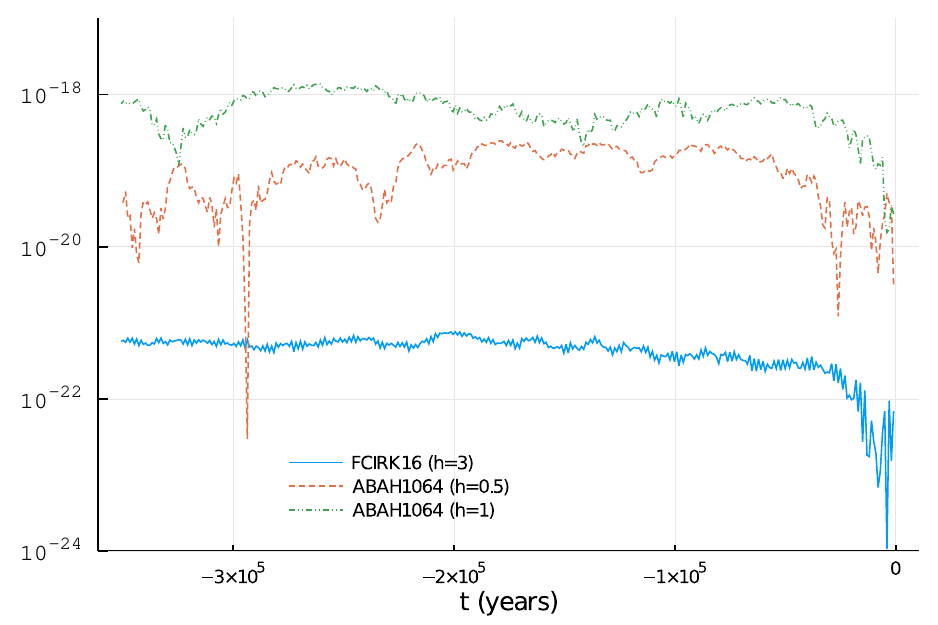}} 
\caption{\small
Evolution of relative errors in energy for FCIRK16 ($h=3$) and ABAH1064 ($h=1$ and $0.5$)
\label{fig:5}}
\end{figure}

 \begin{figure}[ht!]
\centering
\includegraphics[width=.95\textwidth]{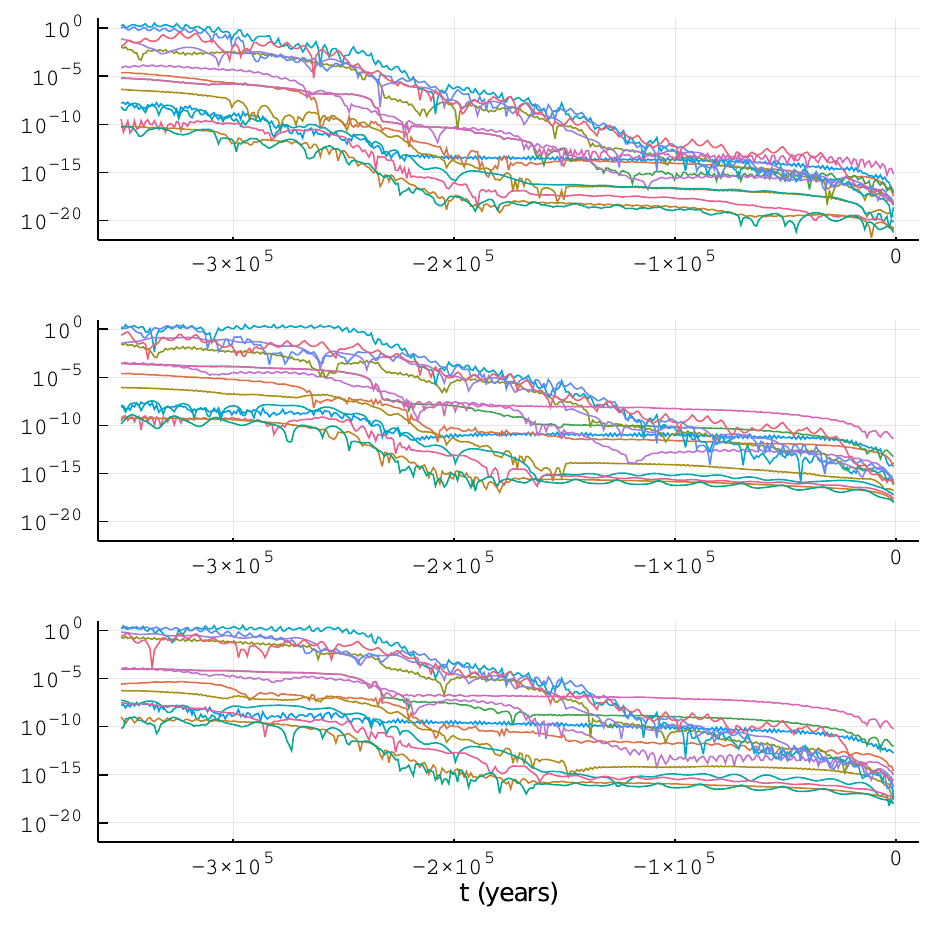}
\caption{\small
Evolution of errors in  position:  FCIRK16 with  $h=3$ (top), and ABAH(10,6,4) with h=0.5 (middle) and
         ABAH(10,6,4) with h=1 (bottom)} 
\label{fig:6}
\end{figure}

In order to illustrate the superiority of FCIRK16 over ABAH(10,6,4) near the limit of precision of the 80-bit floating point arithmetic, we  next compare the propagated errors of three numerical long-term integrations of the 16-body model of the Solar System. We integrate backward in time for 350 thousand years by applying
 \begin{itemize}
 \item FCIRK16 with $h=3$ (16 hours for sequential execution, 8.3 hours in parallel execution with four threads, $0.0015$\% of critical steps).
 \item ABAH(10,6,4) with $h=0.5$  (17 hours of execution time, $0.0019$\% of critical steps).
 \item ABAH(10,6,4) with $h=1$ (8.45 hours of execution time, $0.0019$\% of critical steps). 
 \end{itemize}
 We will next show that the ABAH(10,6,4) integration with $h=1$ (resp. $h=0.5$), which is comparable in computing time to the sequential (resp. parallel) execution of FCIRK, gives less precise results than FCIRK  with $h=3$.
 
\begin{figure}[ht!]
\centering
\includegraphics[width=.95\textwidth]{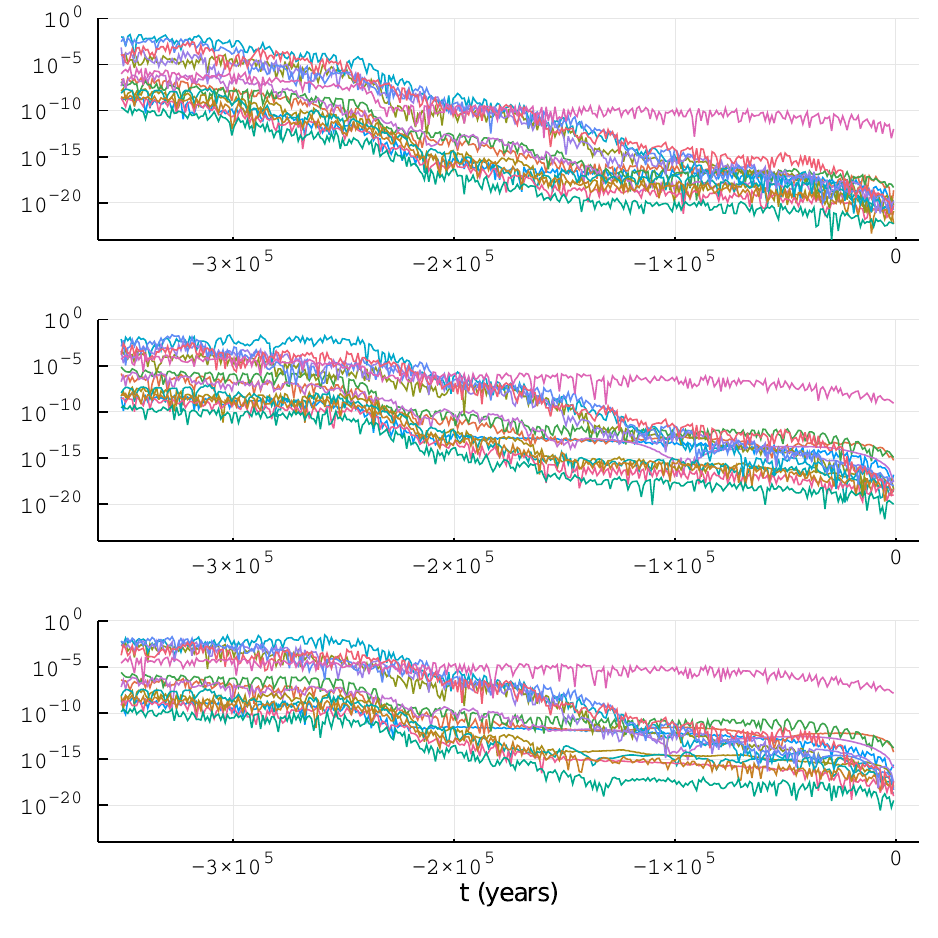}
\caption{\small
Evolution of errors in eccentricity:  FCIRK16 with  $h=3$ (top), and 
ABAH(10,6,4) with h=0.5 (middle) and
         ABAH(10,6,4) with h=1 (bottom)} 
\label{fig:7}
\end{figure}

 We first display the evolution of relative errors in energy for each of the three integrations in Figure~\ref{fig:5}.
Figure~\ref{fig:6} (resp. Figure~\ref{fig:7}) shows the evolution of the errors in position (resp. in eccentricities) of each of the bodies: 
 for FCIRK16 with $h=3$ (top),  for ABAH(10,6,4) with $h=0.5$ (middle), for ABAH(10,6,4) with $h=1$ (bottom).
 The evolution of errors in velocity and in semi-major axis (not included here,)  show a behavior similar to that of positions and eccentricities respectively. 

 Comparison of the evolution of errors for ABAH(10,6,4) with $h=1$ and $h=0.5$, suggests that no substantial increase of the precision could be expected for ABAH(10,6,4) with step-sizes smaller than $h=0.5$. We conclude that for that accuracy regime (limited by the precision of the 80-bit floating point arithmetic), FCIRK is able to produce (about two orders of magnitude) more precise results  than ABAH(10,6,4) for similar execution times. We stress that, as illustrated in Figure~\ref{fig:4}, our implementation of  ABAH(10,6,4) also benefits greatly from the ideas and techniques introduced in Section~\ref{s:close_encounters} and Subsection~\ref{ss:adaptivity}. The evolution of errors for the fully constant step-size implementation of ABAH(10,6,4) would be considerably less favorable due to the occurrence of close encounters between asteroids.

%


\section{Summary of the work}

We have presented FCIRK16, an integrator for high-precision long term integration of the Solar System. It consists of an optimized and robust implementation of  a 16th order scheme within the class of FCIRK methods presented in~\cite{Anto2018}.
Such methods are implicit schemes that take advantage of the near-Keplerian motion of the planets around the Sun  by alternating Keplerian motions with corrections accounting for the planetary interactions. 

We have presented a novel analysis of the local  errors of FCIRK methods intended to understand how discretization errors increase during  close encounters.  That allows us to endow FCIRK16 with a practical mechanisms to monitor and accurately resolve eventual close encounters.

 We have compared our integrator with  a state-of-the-art high precision long term integrator (which is an explicit scheme that also takes advantage of the near-Keplerian motion of the planets), the symplectic splitting method ABAH(10,6,4)~\cite{Blanes2013,Farres2013}. 
 Compared to ABAH(10,6,4) (or any available explicit symplectic integrator), our new integrator FCIRK16 
\begin{itemize}
\item is better suited to parallel implementations, 
\item admits the application of a very efficient mixed precision technique to reduce the accumulation of roundoff errors,  
\item admits any kind of interaction Hamiltonian, not necessarily the sum of integrable parts.
\item   includes a mechanism to detect and effectively cope with eventual close encounters.
\end{itemize}

We present numerical experiments on a point-mass 16-body Newtonian model (including the Moon as a separate body) of the Solar System to compare our integrator  to ABAH(10,6,4) for an integration backward in time of 350 thousand years.  We compare our own optimized implementation in 80-bit extended precision of the ABAH(10,6,4) with a mixed precision implementation (that combines 80-bit extended precision with 128-bit quadruple precision) of FCIRK16. 
Actually, we have compared FCIRK16 with an improved implementation of ABAH(10,6,4) endowed with the same mechanism  to  detect and accurately resolve close encounters.

In our numerical experiments with that 16-body of the Solar System, we conclude that: (i)  ABAH(10,6,4) is more efficient than FCIRK16 for the regime where local truncation errors of ABAH(10,6,4) clearly dominate over the contribution of round-off errors, (ii) for more precise integrations, FCIRK16 is more efficient than ABAH(10,6,4), (iii) the accuracy of ABAH(10,6,4) is limited by the precision of the $80$-bit extended precision floating point arithmetic (and we do not see how to gain some digits of precision without compromising seriously its efficiency), (iv) in that limit, thanks to application of the mixed-precision technique, FCIRK16 is able to improve the precision with similar computing time. We stress that our implementation of ABAH(10,6,4) also benefits from the analysis of Section~\ref{s:close_encounters} and the technique for resolving close encounters proposed in Subsection~\ref{ss:adaptivity}. The superiority for high precisions of FCIRK16 over the standard implementation (with fully constant time-steps) of ABAH(10,6,4) would be considerably higher.

We believe that, for very long integrations such as \cite{Laskar2011}, \cite{Laskar2011b}, the above mentioned extra accuracy may be critical in some cases. For such high precision, it may certainly make sense to use a more realistic model of the Solar System instead of the simple Newtonian point-mass model we have considered in our numerical tests.  That scenario actually favors FCIRK16 over ABAH(10,6,4). Indeed, due to the higher computational cost of the evaluation of $g(u)$ of more realistic models of the Solar System (including expensive relativistic corrections),  (i) the relative parallelization overhead of FCIRK16 should be greatly reduced (hence getting a higher parallelization speedup) compared to the simple model of the Solar System we have considered in our numerical experiments; (ii) in addition, the fixed point implementation could be adapted to perform fewer evaluations of the relativistic corrections per step. This two factors would potentially improve  the relative efficiency of FCIRK16 compared to ABAH(10,6,4). In addition, 
the inclusion of physical effects that cannot be written as the sum of integrable parts may hinder the efficient implementation of ABAH(10,6,4), while this poses no problem to FCIRK16.


FCIRK16 is intended to be applied for planetary systems, like the Solar System, having low eccentricity orbits, and where close encounters between the modeled bodies are relatively rare. For higher eccentricity planetary systems or  systems with potentially frequent close encounters, a fully adaptive version of FCIRK integrators should be implemented.  We believe that the analysis and ideas presented in Section~\ref{s:close_encounters} could be helpful to pursue that goal.

The current C language implementation  of FCIRK16 can be downloaded from our Github software repository: https://github.com/mikelehu/FCIRK/. User documentation is provided and many examples are included, ensuring that the results in the present work are reproducible \cite{Randall2012}.

\section*{Acknowledgements}

All the coauthors have received funding by the Spanish State Research Agency through project PID2019-104927GB-C22, MCIN/AEI/10.13039/501100011033,
and also from the Department of Education of the Basque Government through the Consolidated Research Group MATHMODE (IT1294-19).

\addcontentsline{toc}{chapter}{Bibliography}
\bibliography{bibliography}
\bibliographystyle{ieeetr}


\appendix

\section{Proof of error estimates}

The identity $\varphi_t'(w)^{-1} = \varphi_{-t}'(\varphi_t(w))$  implies that the solution $w(t)$ of (\ref{eq:wODE}) satisfies
 \begin{equation*}
\frac{d}{dt} w(t) = \varphi_{-t}'(u(t))\, g(u(t)),
\end{equation*}
where $u(t)$ is the solution of (\ref{eq:pertODE})--(\ref{eq:icond}) with $u^0=w(0)$.

We denote $R(t,s) := \varphi_{s}'(u(t))\, g(u(t))$, so that 
$\frac{d}{dt}w(t) = R(t,-t)$. 
 Since $\varphi'_t(u)$ is the Jacobian of the $t$-flow $\varphi_t$ of (\ref{eq:keplerODE}), we have that 
\begin{equation}
\label{eq:R(t,s)}
R(t,s) = g(u(t)) + \int_{0}^{s} k'(u(t))\, R(t,\sigma)\, d\sigma,
\end{equation} 
where $k'(u)$ denotes the Jacobian of $k(u)$. 
Here, $k(u)$ and $g(u)$ denote the complex analytic extension of the original real-analytic maps. In particular, each power $\|q_i\|^{\nu} = \|(x,y,z)\|^\nu$ of the Euclidean norm of each 3-vector $q_i = (x_i,y_i,z_i) \in \R^3$ (viz. $\|q_i\|^{-3}$ in (\ref{eq:Kepler_i})) must be replaced by 
$(q_i^T q_i)^{\nu/2} = (x_i^2+y_i^2+z_i^2)^{\nu/2}$,  so that it is analytic for all $(x_i,y_i,z_i)$ such that $x_i^2+y_i^2+z_i^2\neq 0$.

 If we collect $R(t,s)$ in $2n$ three-dimensional sub-vectors as
 $$R(t,s)=(R_1(t,s),\ldots,R_{2n}(t,s))$$
  (so that $R_i(t,0) = g_i(u(t))$)  we have that (\ref{eq:R(t,s)}) is equivalent to the following: 
  for $i=1,\ldots,n$,
\begin{align}
\label{eq:Ri}
R_i(t,s) &= g_i(u(t)) + \int_{0}^{s} R_{i+n}(t,\sigma)\, d\sigma, \\
\label{eq:Rin}
R_{i+n}(t,s) &= g_{i+n}(u(t)) + \int_0^s S_i(t) R_i(t,\sigma)\, d\sigma,
\end{align}
where for arbitrary $t \in \C$, $S_i(t)$ is a $3\times 3$-matrix with complex entries defined as follows: for each $\gamma \in \C^3$,
\begin{equation*}
S_i(t)\gamma := \frac{-k_i}{(q_i(t)^T q_i(t))^{3/2}} \gamma+  \frac32\, (q_i(t)^T \gamma) \frac{k_i}{(q_i(t)^T q_i(t))^{5/2}}
 q_i(t).
\end{equation*}
The following Cauchy estimate then holds,
\begin{align*}
 \frac{1}{(2s)!} \|w_i^{(2s+1)}(0)\|& \leq \rho(u^0)^{-2s} \, \sup_{t \in \mathcal{D}_{\rho(u^0)}} \left\|\frac{d}{dt}w_i(t) \right\| \\
& \leq \rho(u^0)^{-2s} \, \sup_{t,s \in \mathcal{D}_{\rho(u^0)}} \|R_i(t,s)\|.
\end{align*}
From (\ref{eq:Ri})-(\ref{eq:Rin}) we have for $i=1,\ldots,n$,
\begin{equation*}
\|R_i(t,s)\| \leq \|g_i(u(t))\| + |s|\,  \|g_{i+n}(u(t))\| + \frac{|s|^2}{2} \|S_i(t)\|  \|R_i(t,s)\|
\end{equation*}
We will later prove that for all $t \in \mathcal{D}_{\rho(u^0)}$,
\begin{equation}
\label{eq:C(t)ineq}
\rho(u_0)^2 \, \|S_i(t) \gamma\| \leq \|\gamma\|.
\end{equation}
This inequality implies, for $i=1,\ldots,n$,
\begin{equation*}
 \sup_{t,s \in \mathcal{D}_{\rho(u^0)}} \|R_i(t,s)\| \leq 2\, \left( \sup_{t \in \mathcal{D}_{\rho(u^0)}} \|g_i(u(t))\| + \rho(u_0) \, 
  \sup_{t \in \mathcal{D}_{\rho(u^0)}}\|g_{i+n}(u(t))\|  \right),
\end{equation*}
and in turn, taking (\ref{eq:Rin}) into account, 
\begin{equation*}
 \sup_{t,s \in \mathcal{D}_{\rho(u^0)}} \|R_{i+n}(t,s)\| \leq \frac{1}{\rho(u^0)}\, \left( 2\! \sup_{t \in \mathcal{D}_{\rho(u^0)}} \|g_i(u(t))\| + 3\, \rho(u_0) \!
  \sup_{t \in \mathcal{D}_{\rho(u^0)}}\|g_{i+n}(u(t))\|  \right).
\end{equation*}

It then only remains to prove the inequality (\ref{eq:C(t)ineq}):
Condition C2 implies (with the help of Lemma~1 in~\cite{Anto2020}) that,  for all $t \in \mathcal{D}_{\rho(u^0)}$ and all $\gamma \in \C^3$,
\begin{equation}
\label{eq:Caux1}
\|S_i(t) \gamma\| \leq \xi(1/7)\, \frac{k_i}{\|q_i^0\|^3}\, \|\gamma\|,
\end{equation}
where
\begin{equation*}
\xi(\lambda) =  \frac{1}{(1-2\lambda -\lambda^2)^{3/2}} + \frac32\, \frac{(1+\lambda)^2}{(1-2\lambda -\lambda^2)^{5/2}}.
\end{equation*}
From the other hand, 
\begin{equation}
\label{eq:Caux2}
\rho(u^0)^2 \leq \frac17\, \frac{\|Q_i^0-Q_0^0\|}{K_i(Q)+K_0(Q)} 
                   \leq \frac17\, \frac{\|Q_i^0-Q_0^0\|^3}{G\, (m_0 + m_i)} 
                   = \frac17\, \frac{\|q_0^0\|^3}{k_i}.
\end{equation}
The inequality (\ref{eq:C(t)ineq}) follows from (\ref{eq:Caux1})--(\ref{eq:Caux2}) as $\frac{1}{7}\xi(1/7) = 0.945022\ldots < 1$.

\end{document}